\documentclass[12pt,a4paper]{elsart}

\usepackage{epsfig}
\usepackage{amsmath}
\usepackage{graphics}
\usepackage{amssymb}

\begin{document}

\begin{frontmatter}



\title{Optimization of network structure to random failures}

% \author{author\corauthref{cor1}\thanksref{label2}}

\author[Bing Wang]{Bing Wang\corauthref{cor}} \corauth[cor]{Corresponding author.}\ead{bingbignmath@yahoo.com.cn }
\author[Bing Wang] {Huanwen Tang}
 \ead{imsetang@dlut.edu.cn}
\author[Bing Wang]{ Chonghui Guo}\ead{guochonghui@tsinghua.org.cn}
\address[Bing Wang]{Department of Applied Mathematics, Dalian University of Technology, Dalian Liaoning, 116024, P. R. China}
\author[Xiu]{Zhilong Xiu}
\address[Xiu]{School of Environmental and Biological Science and Technology, Dalian University of
Technology, Dalian Liaoning, 116024, P. R. China}
\ead{zhlxiu@dlut.edu.cn}
\author[Zhou]{Tao Zhou}
\address[Zhou]{Nonlinear Science Center and Department of Modern Physics, University of Science and Technology of China,
Hefei Anhui, 230026, P. R. China}
\ead{zhutou@ustc.edu}

%\title{}
%\author{}
\begin{abstract}
Network's resilience to the malfunction of its components has been
of great concern. The goal of this work is to determine the
network design guidelines, which maximizes the network efficiency
while keeping the cost of the network ( that is the average
connectivity ) constant. With a global optimization method, memory
tabu search ( MTS ), we get the optimal network structure with the
approximately best efficiency. We analyze the statistical
characters of the network and find that a network with a small
quantity of hub nodes, high degree of clustering may be much more
resilient to perturbations than a random network and the optimal
network is one kind of highly heterogeneous networks. The results
strongly suggest that networks with higher efficiency are more
robust to random failures. In addition, we propose a simple model
to describe the statistical properties of the optimal network and
investigate the synchronizability of this model.
\end{abstract}

\end{frontmatter}

% -------------------- Introduction --------------------
\section{Introduction}
Complex networks arisen in natural and manmade systems play an
essential role in modern society. Many real complex networks were
found to be heterogeneous with power-law degree distributions:
$p(k)\sim k^{-\gamma}$, such as the Internet, metabolic networks,
scientific citation networks, and so on
\cite{Albert02,Strogatz01}. Because of the ubiquity of scale-free
networks in natural and manmade systems, the security of these
networks, i.e., how well these networks work under failures or
attacks, has been of great concern.

Recently, a great deal of attention has been devoted to the
analysis of error and attack resilience of both artificially
generated topologies and real world networks
\cite{Albert00,Cohen00,Cohen01,Crucitti04,Crucitti042,Holme02,Gallos04,Magoni03,Zhou05}.
Also some researchers use the optimization approaches to improve
the network's robustness with percolation theory or information
theory \cite{Valente04,Tanizawa05,Paul04,Callaway00,Wang05,Liu05}.
There are various ways in which nodes and links can be removed,
and different networks exhibit diverse levels of resilience to
such disturbances. It has been pointed out by a number of authors
\cite{Albert00,Cohen00,Cohen01,Crucitti04,Crucitti042} that
scale-free networks are resilient to random failures, while
fragile to intentional attacks. That is, intentional attack on the
largest degree (or betweeness) node will increase the average
shortest path length greatly. While random networks show similar
performance to random failures and intentional attacks.

The network robustness is usually measured by the average
node-node distance, the size of the largest connected subgraph, or
the average inverse geodestic length named \emph{efficiency} as a
function of the percentage of nodes removed. Efficiency has been
introduced in the studies of small world networks \cite{Latora01}
and used to evaluate how well a system works before and after the
removal of a set of nodes \cite{Crucitti04}.

The network structure and function strongly rely on the existence
of paths between pairs of nodes. Different connectivity pattern
between pairs of nodes makes the network different performance to
attacks. Rewiring edges between different nodes to change the
topological structure may improve the network's function. As an
example, consider the simple five nodes network shown in Fig. 1.
The efficiency of Fig. 1(a) is equal to 8/25, while it is improved
to 7/20 in Fig. 1(b) by rewiring. And we know that Fig. 1(b) is
more robust than Fig. 1(a) to random failures. A natural question
is addressed: how to optimize the robustness of a network when the
cost of the network is given. That is, the number of links remains
constant while the nodes connect in a different way. Should the
network have any particular statistical characters? This question
motivates us to use a heuristic approach to optimize the network's
function by changing the network structure.

The paper is organized as follows: we firstly present MTS method
in Section 2 and the numerical results are shown in Section 3.
Then we construct a simple model to describe the optimal network
and discuss one of the important dynamic processes happening on
the network, synchronization, in Section 4. Finally, we give some
insightful indications in Section 5.
\begin{center}
\begin{figure}
\centerline{\psfig{file=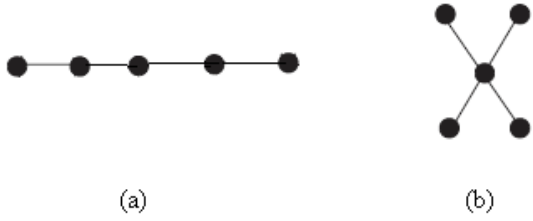,width=7cm}} \vspace*{8pt}
\caption{(a) $\langle{k}\rangle=8/5, E=8/25$; (b)
$\langle{k}\rangle=8/5, E=7/20$.}
\end{figure}
\end{center}
%section 2
% -------------------- section 2. ----------------------
\section{The algorithm}
Generally, a network can be described as an unweighted, undirected
graph $G$. Such a graph can be presented by an adjacency binary
matrix $A=\{a_{ij}\}$. $a_{ij}=1$ if and only if there is an edge
between node $i$ and $j$. Another concerned matrix $D=\{d_{ij}\}$,
named distance matrix, consists of the elements denoting the
shortest path length between any two different nodes. Then the
efficiency $\varepsilon_{ij}$ between nodes $i$ and $j$ can be
defined to be inversely proportional to the shortest distance:
 $\varepsilon_{ij}=1/d_{ij}$ \cite{Latora01}.
 The global efficiency of the network is
defined as the average of the efficiency over all couples of
nodes.
\begin{center}
\begin{equation}
E(G)=\frac{1}{N(N-1)}\sum_{i\neq{j}\in{G}}{{\varepsilon_{ij}}}=\frac{1}{N(N-1)}\sum_{i\neq{j}\in{G}}{\frac{1}{d_{ij}}}
\end{equation}
\end{center}
With the above robustness criterion in mind, we can define the
optimization problem as follows:
\begin{center}
\begin{eqnarray}
 \left\{ {{\begin{array}{l}
{max E(G)}\\
s.t. {\langle{k}\rangle}={const}\\
 {\ \ \ \ G \ is \ connected.}\\
\end{array} }} \right.
\end{eqnarray}
\end{center}
The above problem is a standard combinatorial optimization
problem, for which we can derive good ( though usually not
perfect) solutions using one of the heuristic algorithms, tabu
search, which is based on memory ( MTS ) \cite{Ji04}. MTS is
described as follows:

$step\ 1$: Generate an initial random graph $G_{0}$ with $N$
nodes, $M$ edges. Set $G_{t}^{*}:=G_{0}$; $t:=0$. Compute the
efficiency $E$ of $G_{t}^{*}$ denoted by $E_{G_{t}^{*}}$.

$step\ 2$: If a prescribed terminal condition is satisfied, stop,
otherwise random rewiring: specifically, a link connecting node
$i$ and $j$ is randomly chosen and substituted with a link from
$i$ to node $k$, not already connected to $i$, extracted with
uniform probability among the nodes of the network, note the
present network $G$ and the efficiency $E_{G}$.

$step\ 3$: If $E_{G}\geq{E_{G_{t}^{*}}}$, $G_{t+1}^{*}:=G$,
$G_{t+1}:=G$, else if $E_{G}\geq{E_{G_{t}}}$, $G_{t+1}:=G$, else
if $G$ does not satisfy the tabu conditions, then $G_{t+1}:=G$,
else $G_{t+1}:=G_{t}$. Go to $step\ 2$.

The following condition is used to determine if a move is tabu:
$\frac{|E_{G_{t}}-E_{G}|}{E_{G}}>\delta$, which is the percentage
improvement or destruction that will be accepted if the new move
is accepted. Thus, the new graph at $step\ 2$ is assumed tabu if
the total change in the objective function is higher than a
percentage $\delta$. In this paper, $\delta$ is a random number
generated between 0.50 and 0.75. The terminal condition is that
the present step is getting to the predefined maximal iteration
steps.
\section{Numerical results}
Many real networks in nature and society share two generic
properties: scale-free degree distribution and small-world effect
( high clustering and short path length). Another important
property of a network is the degree correlation of node $i$ and
its neighbors. It is called \emph{assortative} mixing if
high-degree nodes are preferentially connected with other
high-degree nodes, and \emph{disassortative} mixing if high-degree
nodes attach to low-degree nodes. Newman proposed a simple measure
to describe the mixing pattern of nodes, which is a correlation
function of the degrees. The empirically studied results show that
almost all the social networks show assortive mixing pattern while
other technological and biological networks are disassortative.
The statistical properties are clearly described in Refs
\cite{Zhang05,Amaral00,Newman02,Newman04}. We start from a random
graph with size $N=300$ and $\langle{k}\rangle=6.7$. The terminal
condition is the maximal iteration step reaching 1000. It should
be noted that for each step of the objective function being
improved, we record the statistical properties of the present
network. A typical run of statistical results are shown in Figs. 2
and 3.
%random results fig1
\begin{center}
\begin{figure}
\centerline{\psfig{file=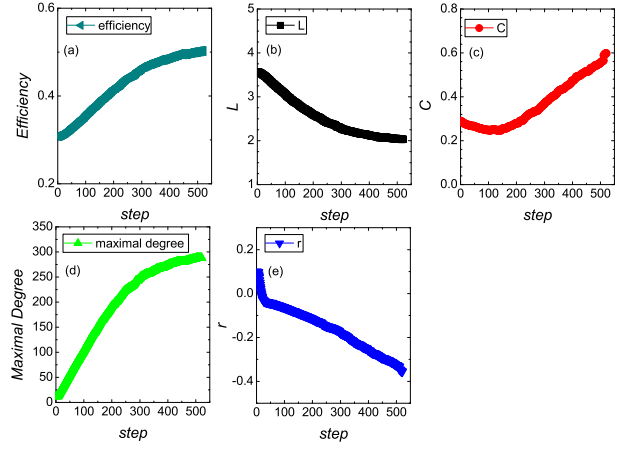,width=9cm}} \vspace*{8pt}
\caption{The characteristics of the optimal network evolving from
a random network. (a) Efficiency of the network; (b) The average
path length $L$; (c) The average clustering coefficient $C$; (d)
The maximal degree of the network; (e) The degree correlation
coefficient $r$. The network size is $N=300$ and
$\langle{k}\rangle=6.7$.}
\end{figure}
\end{center}
Fig. 2 shows that with the increase of efficiency $E$, the average
shortest path length $L$ becomes short and the maximal degree
becomes large, indicating that hub nodes develop to be present
with the evolving process. With the increase of the efficiency,
the hub node develops to be the most important one to connect with
almost other nodes in the network. For degree correlation
coefficient $r$ (Fig2. (e)), it decreases in the whole process
from zero to negative, which indicates that the nodes with high
degree preferentially connect with the ones with low degree.
Still, for the clustering coefficient $C$, it increases to a high
value 0.6 and the network gets to be a highly clustering network.
For the degree distribution, the cumulative degree distribution is
shown in Fig. 3.

To check the optimal network's tolerance to random failures, we
show the efficiencies of both initial random network and the
optimal network versus the fraction of removed nodes in Fig. 4. It
can be clearly observed that compared with the initial random
network, the optimal network's robustness to errors is greatly
improved.
\begin{center}
\begin{figure}
\centerline{\psfig{file=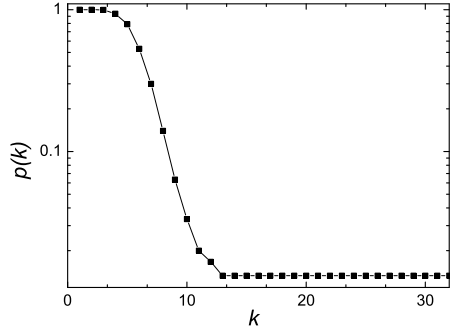,width=7cm}} \vspace*{8pt}
\caption{The cumulative degree distribution of the optimal network
in log-linear scale. The network size is $N=300$ and
$\langle{k}\rangle=6.7$.}
\end{figure}
\end{center}
\begin{center}
\begin{figure}
\centerline{\psfig{file=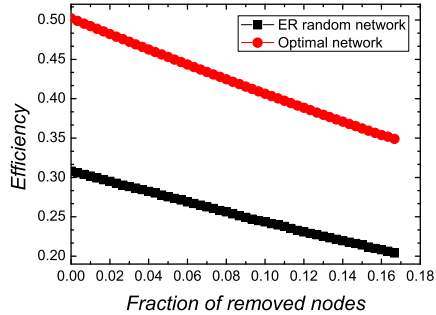,width=7cm}} \vspace*{8pt}
\caption{Efficiencies of the optimal network and the initial
random network versus the fraction of removed nodes. The data are
averaged over 10 independent runs of network size $N=300$ and
$\langle{k}\rangle=6.7$.}
\end{figure}
\end{center}
\section{Model and synchronization of the network}
To provide a simple way to describe the properties of the optimal
network, we consider to construct the network model directly. With
a nongrowing network model it can be constructed in the following
way.

(a) Start from a random network with $N$ nodes, which can be
implemented by rewiring edges of a regular graph with probability
$p=1$.

(b) Choose $q$ nodes as hub nodes randomly from the whole network
with equal probability.

(c) Add $m$ edges randomly. One end point of the edge is selected
randomly from the $q$ hub nodes and the other is chosen randomly
from the network.

In such a way, the network evolves to possess the statistical
properties of the optimal network. This can be seen from Fig. 5.
The primary goal of our simulation is to understand how the
statistical properties of the network change with the process of
adding edges. The construction of the model is similar to the
two-layer model introduced by Nishikawa \emph{et. al} in Ref
\cite{Nishikawa03}. The main difference is that the initial
network in our model is a random network different from that of a
regular network.
%----figure5----
\begin{center}
\begin{figure}
\centerline{\psfig{file=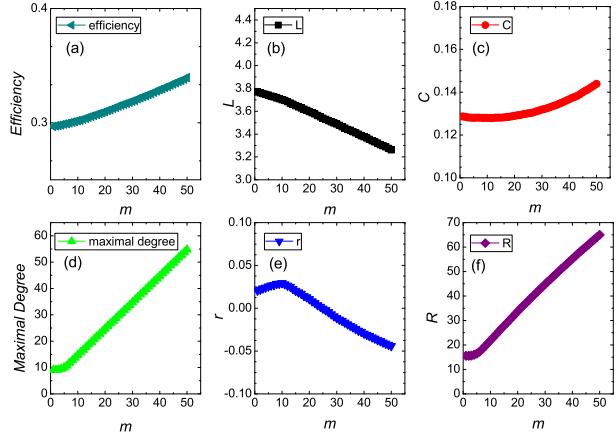, width=9cm}} \vspace*{8pt}
\caption{The statistical properties of the network model versus
the parameter $m$. (a) Efficiency of the network; (b) The average
path length $L$; (c) The average clustering coefficient $C$; (d)
The maximal degree of the network; (e) The degree correlation
coefficient $r$; (f) The eigenvalue ratio $R$ of the network,
which is an important measure of network synchronizability.  The
network size is $N=200, q=1, \langle{k}\rangle=5.2.$ All data are
averaged over 100 realizations.}
\end{figure}
\end{center}

To show the effect of the parameter $q$, we also present
simulation results versus $q$ in Fig. 6. With the increase of the
parameter $q$, the network becomes less heterogeneous and more
homogeneous, it's natural to observe that both the efficiency of
the network and the maximal degree reduce. So both the average
path length $L$ and the correlation coefficient $r$ increase in
the homogeneous phase compared with the values in the
heterogeneous phase. Since the optimal network shows a strong
heterogeneity, a small parameter of $q$ is reasonable. We know
that most of the real world networks share the character of
small-world effect and some degree of heterogeneity, so these
networks are robust to random failures and they are also efficient
in exchanging information.
%----figure 6----
\begin{center}
\begin{figure}
 \centerline{\psfig{file=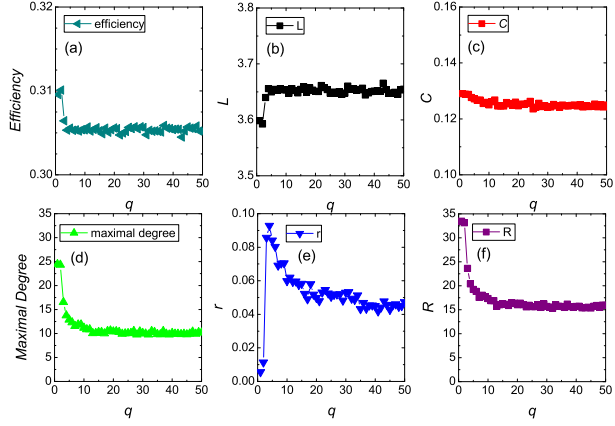,width=9cm}} \vspace*{8pt}
\caption{The statistical properties of the network model versus
the parameter $q$. (a) Efficiency of the network; (b) The average
path length $L$; (c) The average clustering coefficient $C$; (d)
The maximal degree of the network; (e) The degree correlation
coefficient $r$; (f) The eigenvalue ratio $R$ of the network,
which is an important measure of network synchronizability. The
network size is $N=200, m=30, \langle{k}\rangle=6$. In all
simulations, 100 realizations are averaged.}
\end{figure}
\end{center}

Then we consider the synchronization of the network model, how
does the network's synchronizability change with the adding of
edges? Synchronization has been observed in diverse natural,
social and biological systems. Consider a network consisting of
$N$ identical oscillators coupled through the edges of the
network. The dynamics of each individual oscillator is controlled
by $\dot{\textbf{x}}_i=\textbf{f}(\textbf{x}_i)$ and
$\textbf{h}(x_j)$ is the output function. Thus, the equations of
motion are as follows:
\begin{equation}
\dot{\textbf{x}}_i=\textbf{f}(\textbf{x}_i)-\sigma\sum_{j=1}^NL_{ij}\textbf{h}(\textbf{x}_j),
\end{equation}
where $\dot{\textbf{x}}_i=\textbf{f}(\textbf{x}_i)$ governs the
dynamics of individual oscillator, $\sigma$ is the coupling
strength, and $L=\{L_{ij}\}$ is the Laplacian matrix of the
network. It has been shown that the eigenvalue ratio
$R=\frac{\lambda_{N}}{\lambda_{2}}$ is an essential measure of the
network synchronizability, the smaller the eigenvalue ratio $R$,
the easier the network to synchronize \cite{Barahona02}. The
progress in the studies of the relationship between topological
structure and synchronizability can be found in Refs
\cite{wang05b,Donetti05,Zhou05a,Nishikawa03,Zhao05,Hong04,Bernardo05,McGraw05}.

To discuss the synchronizability of the network model, we show the
eigenvalue ratio $R$ versus the number of adding edges $m$ and the
number of adding hub nodes $q$ in Fig. 5 (f) and Fig. 6 (f)
respectively. Note that in Fig. 5 (f), the case for $q=1$
corresponds to the highest heterogeneity. The eigenvalue ratio $R$
increases with $m$ greatly, which means that the network becomes
more difficult to synchronize with strong heterogeneity, even for
short path length. In Fig. 6 (f), the network synchronizability is
improved with the increase of $q$. These can all be explained as
strong heterogeneity reduces network's synchronizability, which is
consistent with the conclusion of Nishikawa \emph{et. al} who has
pointed out that networks with homogeneous distributions are more
synchronizable than heterogeneous ones \cite{Nishikawa03}.

We can conclude that with the introduction of heterogeneity,
though the network robustness to random failures and the
efficiency of information exchange on the network are greatly
improved, the network's synchronizability is really reduced.
\section{Conclusions}
One ultimate goal of studies on complex networks is to understand
the relationship between the network structure and its functions.
To get the optimal strategies of a given function, we should
evolve the structure with its function dynamically, which can be
realized with optimization approaches. What special characters
should the network have with a given function? This problem
motivates us to explore the relationship between the structure
properties and functions and then get some insightful conclusions.

By optimizing the network structure to improve the performance of
the network resilience, we obtain the optimal network and do some
statistics of the optimal network. We find that during the
optimizing process, the average shortest path length $L$ becomes
short. The increase of the maximal degree of the network indicates
the hub nodes' appearance. The degree correlation coefficient $r$
decreases and is always less than zero, which indicates that nodes
with high degree preferentially connect with the low degree ones.
The clustering coefficient $C$ increases in the whole process and
arrives to a high level, then the network shows a high degree of
clustering. As we all know that most of the real-world networks in
social networks show high clustering, short path length and
heterogeneity of degree distributions, which may indicate their
good performance to random failures and high efficiency of
information exchange. Then we present a nongrowing network model
to try to describe the statistical properties of the optimal
network and also analyze the synchronizability of the network. And
we find that although the network's robustness to random failures
and the efficiency of information exchange are greatly improved
(for the average distance of the network is small), the network's
synchronizability is really reduced for the network's strong
heterogeneity.

In summary, we try an alternative point of view to analyze the
robustness of the network from its efficiency. By optimizing the
network efficiency we find that a network with a small quantity of
hub nodes, high degree of clustering may be much more resilient to
perturbations. And the results strongly suggest that the network
with higher efficiency are more robust to random failures, though
its synchronizability is being reduced greatly.

%------------Appendix---------

% ------------- References -----------------------------

\end{document}